**MIFA: Metadata, Incentives, Formats, and Accessibility guidelines to improve the reuse of AI datasets for bioimage analysis**

Teresa Zulueta-Coarasa[1], Florian Jug[2], Aastha Mathur[3], Josh Moore[4], Arrate Muñoz-Barrutia[5,6], Liviu Anita[1], Kolawole Babalola[1], Peter Bankhead[7], Perrine Gilloteaux[8], Nodar Gogoberidze[9], Martin L. Jones[10], Gerard J. Kleywegt[1], Paul Korir[1], Anna Kreshuk[11], Aybüke Küpcü Yoldaş[1], Luca Marconato[12,13,14], Kedar Narayan[15,16], Nils Norlin[17], Bugra Oezdemir[3], Jessica L. Riesterer[18], Craig Russell[1], Norman Rzepka[19], Ugis Sarkans[1], Beatriz Serrano-Solano[3], Christian Tischer[20], Virginie Uhlmann[1,21], Vladimír Ulman[22,23], and Matthew Hartley[1]

[1] European Molecular Biology Laboratory, European Bioinformatics Institute, Hinxton, UK

[2] Fondazione Human Technopole, V.le Rita Levi-Montalcini, 1, 20157 Milan, Italy

[3] Euro-BioImaging ERIC Bio-Hub, European Molecular Biology Laboratory (EMBL) Heidelberg, Meyerhofstraße 1, 69117 Heidelberg, Germany

[4] German BioImaging – Gesellschaft für Mikroskopie und Bildanalyse e.V., Konstanz, Germany

[5] Universidad Carlos III de Madrid, Leganés, Madrid, Spain

[6] Instituto de Investigación Sanitaria Gregorio Marañón, Madrid, Spain

[7] Edinburgh Pathology, Centre for Genomic and Experimental Medicine, and CRUK Scotland Centre, Institute of Genetics and Cancer, University of Edinburgh, Edinburgh, UK

[8] Nantes Université, CHU Nantes, CNRS, Inserm, BioCore, US16, SFR Bonamy, F 44000 Nantes, France

[9] Imaging Platform, Broad Institute, Cambridge, United States

[10] Electron Microscopy Science Technology Platform, Francis Crick Institute, London, UK

[11] Cell Biology and Biophysics Unit, European Molecular Biology Laboratory, 69117, Heidelberg, Germany

[12] European Molecular Biology Laboratory (EMBL), Genome Biology Unit, Heidelberg, Germany

[13] Division of Computational Genomics and System Genetics, German Cancer Research Center (DKFZ), Heidelberg, Germany

[14] Collaboration for joint PhD degree between EMBL and Heidelberg University, Faculty of Biosciences

[15] Center for Molecular Microscopy, Center for Cancer Research, National Cancer Institute, NIH, Bethesda, Maryland, USA
[16] Cancer Research Technology Program, Frederick National Laboratory for Cancer Research, Frederick, Maryland, USA

[17] Department of Experimental Medical Science, Lund University Bioimaging Centre (LBIC), & Nanolund, Lund University, Lund, Sweden

[18] Cancer Early Detection Advanced Research (CEDAR), Knight Cancer Institute, Oregon Health & Science University, Portland, OR, USA

[19] Scalable Minds GmbH, Potsdam, Germany

[20] European Molecular Biology Laboratory, Data Science, Heidelberg, Germany

[21] BioVisionCenter, Universität Zürich, Zürich, Switzerland

[22] IT4Innovations, VSB – Technical University of Ostrava, Ostrava-Poruba, Czech Republic

[23] CEITEC - Central European Institute of Technology, Masaryk University, Brno, Czech Republic

## Abstract

Artificial Intelligence methods are powerful tools for biological image analysis and processing. High-quality annotated images are key to training and developing new algorithms, but access to such data is often hindered by the lack of standards for sharing datasets. We discuss the barriers to sharing annotated image datasets and suggest specific guidelines to improve the reuse of bioimages and annotations for AI applications. These include standards on data formats, metadata, data presentation and sharing, and incentives to generate new datasets. We are positive that the MIFA (Metadata, Incentives, Formats, and Accessibility) recommendations will accelerate the development of AI tools for bioimage analysis by facilitating access to high-quality training and benchmarking data.

## Introduction

Imaging is an essential tool in molecular, cell, and developmental biology. Advances in microscopy have endowed scientists with the ability to investigate biological processes across different scales, from imaging molecules at unprecedented resolution (Zhang et al. 2020; Nakane et al. 2020; Yip et al. 2020; Betzig et al. 2006; Rust, Bates, and Zhuang 2006; Hess, Girirajan, and Mason 2006), to recording whole organisms over time (Megason 2009; McDole et al. 2018; Daetwyler et al. 2019; Chen et al. 2014; Huisken et al. 2004; Udan et al. 2014; Royer et al. 2016). Such microscopy experiments can generate image data amounting to TBs in size and, in many cases, quantitative automated analysis must be undertaken to extract meaningful conclusions. Artificial Intelligence (AI), Machine Learning (ML, Supplementary Box 1, (Moen et al. 2019)), and Deep Learning (DL) methods have emerged as essential tools for automated bioimage analysis (Moen et al. 2019; Hallou et al. 2021;

Gupta et al. 2019; Villoutreix 2021; Wang et al. 2019). DL models are capable of identifying the most representative features for specific image-related tasks (e.g., segmentation). They utilise these features to convert input images into desired outputs, such as segmentation masks, while simultaneously learning more complex features (LeCun, Bengio, and Hinton 2015). These features can often be learned on a subset of the data and then shown to generalise well to the entire dataset. Therefore, DL is particularly suitable for analysing large microscopy datasets while minimising human interaction.

The two most common paradigms in ML are supervised and unsupervised learning (Villoutreix 2021). Supervised approaches take advantage of human knowledge in the form of a ground truth annotated training set to learn the relationship between input data and the desired output. Examples of annotations include class labels, bounding boxes, or segmentation masks (Box 1, Figure 1). On the other hand, unsupervised approaches find underlying patterns in input data without being trained with exemplary outputs. Given that currently the most successful ML models for bioimage analysis are supervised, having access to annotated datasets according to the FAIR (Findability, Accessibility, Interoperability, and Reusability) principles (Wilkinson et al. 2016) is crucial for model development. Furthermore, most DL models are domain-specific, and their performance depends on the data on which the model is trained (Rutschi, Berente, and Nwanganga 2023). A strategy to cover a larger domain and to make models more reusable and generalizable is to use large, heterogeneous, annotated datasets for training (Laine et al. 2021).

**Box 1. Annotation types**

**Bounding boxes:** rectangles or rectangular prisms completely enclosing a structure of interest within an image.

**Class labels:** tags that identify specific features, patterns, or classes in images. They can be given for a whole image or for individual structures within it.

**Counts:** number of objects, such as cells, found in an image.

**Derived annotations:** additional analytical data extracted from the images. For example, the image point spread function, the signal-to-noise ratio, focus information, etc.

**Geometrical annotations:** polygons and shapes that follow the outline of a region of interest in the image. These can be geometrical primitives, 2D polygons, 3D meshes, etc.

**Graphs:** graphical representations of the morphology, connectivity, or spatial arrangement of biological structures in an image. Graphs, such as skeletons or connectivity diagrams, typically consist of nodes and edges, where nodes represent individual elements or regions and edges represent the connections or interactions between them.

**Point annotations:** X, Y, and Z coordinates of a point of interest in an image (for example, an object's centroid or biological landmarks such as particle positions in cryoEM, cryoET, or spatial transcriptomics).

**Segmentation masks:** an image, the same size as the source image, with the value of each pixel representing some biological identity or background region.

**Tracks:** annotations marking the movement or trajectory of objects within a sequence of bioimages.

Despite how crucial training sets are for model training, testing, validation and benchmarking, access to high-quality curated images is scarce. Generating well-annotated datasets is time-consuming, requiring experts to manually annotate on the order of hundreds to thousands of images or to curate machine-created annotations. Therefore, the quantity of datasets to train AI models, the number of annotated images, and the density of annotations within a dataset can be insufficient for a model to generalise effectively beyond the training set. Additionally, due to the lack of a central repository for curated AI-ready annotated datasets, the training sets that are published are scattered in different storage locations, hindering their findability (Laine et al. 2021). This also results in datasets being published using diverse formats, with varying degrees of metadata and unclear licenses, making reuse difficult. As scientists from different fields use different vocabularies to describe their data, dataset reuse becomes challenging when model developers lack a proper understanding of the biological domain, highlighting the importance of clear metadata.

Many community efforts have been put in place to improve dataset development, annotation, general metadata standardisation, and data findability. Consortia such as QUAREP-LiMi (Boehm et al. 2021) or the Open Microscopy Environment (OME) work to provide standards for bioimage data (Swedlow et al. 2021) and metadata (Linkert et al. 2010) formats. Furthermore, metadata standards have emerged such as the REcommended Metadata for Biological Images (REMBI) guidelines for microscopy images from multiple modalities (Sarkans et al. 2021), and the Minimum Information about Highly Multiplexed Tissue Imaging (MITI) standard, which provides guidelines for multiplexed and histology images (Schapiro et al. 2022). Valuable large annotated datasets and datasets curated specifically for AI purposes have been recently made available (Schwendy, Unger, and Parekh 2020; Stringer et al. 2021; Edlund et al. 2021; Conrad and Narayan 2021, 2023). In addition, useful datasets have been published as part of competitions such as the 2018 Data Science Bowl (Caicedo et al. 2019) or the Cell Tracking Challenge (Ulman et al. 2017; Maška et al. 2023), and in resources like the Broad Bioimage Benchmark Collection (Ljosa, Sokolnicki, and Carpenter 2012), the Electron Microscopy Public Image Archive, (Iudin et al. 2023) or the BioImage Archive (Hartley et al. 2022). The “papers with code” initiative hosts datasets related to ML publications (https://paperswithcode.com/datasets). Additionally, journals such as Data in Brief (https://www.sciencedirect.com/journal/data-in-brief) and Scientific Data (https://www.nature.com/sdata/) focus on publishing articles that provide access to research data, crediting the data owners. However, to the best of our knowledge, there are no largely adopted guidelines for bioimage annotation metadata and formats in the field. Developers of AI methods for bioimage analysis would strongly benefit from a specialised repository that provides easy access to images with the corresponding standardised annotations and metadata.

To improve accessibility and re-usability of annotated image datasets, we describe a series of recommendations on four main topics: Metadata, Incentives, Formats, and Accessibility (MIFA). Metadata is essential to enable data re-use and search within repositories, in the first section we present a standard for metadata for biological image annotations. Then we discuss ways to incentivise the production of AI-ready datasets by both crediting annotators and organising events focused on data annotation and dataset curation. The third section considers different annotation file formats that address key accessibility considerations, such as being cloud-ready to support analysis of large datasets without download, handling multi-dimensional data, and being compatible with different community tools. Finally, we examine ways to improve data presentation, and retrieval from repositories.

**The devil is in the metadata: towards a recommended metadata standard**

Standardising metadata enhances the findability and reusability of data. We recommend including four main categories of accompanying metadata to maximise the reuse of an AI dataset (Figure 2). These include general study metadata (providing experimental and investigative context), metadata related to the images, metadata related to the annotations, and metadata related to versioning. Supplementary Table 1 contains metadata entries as well as two examples.

The **study-level metadata** must include a brief description detailing the biological application of the dataset to facilitate the reuse of data by researchers from different fields. These metadata must also identify the authors and publications related to the dataset. Furthermore, to increase findability and interoperability, persistent identifiers, ontologies and controlled vocabularies (Bard and Rhee 2004) should be used when possible. The lack of clear copyright permissions is one of the main impediments developers encounter when reutilizing datasets. Therefore, the type of license the images and the annotations are under must be clearly specified, and open licenses that encourage reuse, such as CC0 ("CC0 - Creative Commons" 2009) and CC BY ("Creative Commons — Attribution 4.0 International — CC BY 4.0," n.d.) are preferred. Finally, the study-level metadata should include pointers to any AI models trained using the dataset stored in specialised repositories such as the BioImage Model Zoo (Ouyang et al. 2022) and vice versa.

The REcommended Metadata for Biological Images (REMBI) guidelines (Sarkans et al. 2021) should be used for the dataset **images metadata**, covering information about the biological material used in imaging - organisms and sample preparation, together with technical details of imaging methods and image analysis among others.

The attributes of the **annotations metadata** module are summarised in Table 1.

| **Attribute** | **Comments** |
| --- | --- |
| Authors and contact | People involved in creating or curating annotations. Include contact information, such as email, ORCiD, GitHub account, or Google Scholar link. |
| Annotation overview | Short description of the annotation and how it was generated. |
| Annotation type | Annotation type, for example, class labels, segmentation mask, or object counts (Box 1). |
| Annotation method | Crowdsourced or expertly annotated. Produced by a human or software (e.g., synthetic data or silver-standard annotations obtained by combining the results of benchmarked algorithms (Maška et al. 2023)). Software used and protocols used for consensus and quality assurance. |
| Annotation confidence level | Confidence in annotation accuracy (e.g., self-reported confidence, the variance among several annotators (Maška et al. 2023), or the number of years of experience of the annotator generating those particular annotations). It can also be noted here if these are weak annotations: rough, imprecise annotations that are fast to generate. Weak annotations are used, for example, to detect an object without providing accurate boundaries. |
| Annotation criteria | Rules used to generate annotations. For example, when counting cells in an image, at what point a dividing cell is considered two different cells? |
| Annotation coverage | Which images or ROIs from the dataset were annotated, and what percentage of the data has been annotated from what is available? If any images or ROIs have not been annotated, what are the reasons? |
| Source image association | Association between annotations and the source images from which they originated. If the annotation refers to a dataset separate from that containing the annotations, it should include the unique identifier for that dataset. |
| Transformations | Any coordinate transformation required to link the images to the annotations in the same common coordinate system, such as rotations and translations. |
| Spatial information | Spatial information for non-pixel annotations (e.g., counts of items in a region of interest), including physical measurements or the region that has been annotated. If coordinates for non-rasterized annotations are stored in units other than pixels (e.g. µm), the relevant information must be detailed in this field. |
| Last modification time | Date and time when the annotation was last modified. |

**Table 1. Annotations metadata module.**

The final module is related to **versioning metadata**. Incremental versions of the dataset must have accompanying metadata with timestamps and a pointer to the previous version. A textual description of the changes between versions should also be included and, if a new version is based on a previous one, the new version should preserve the credit to the original annotators. In all cases, the identity of the creators of the original images, licensing of those images, and a link to the original images must be maintained.

**Credit where credit's due: incentivising production and sharing of AI-ready datasets**

Encouraging sharing annotated image datasets through open-access data repositories requires both practical recognition of those sharing data by the scientific community, for example as part of researcher assessment, and mechanisms to support the attribution of shared datasets to individuals and groups.

One measure to support recognition of the contribution of annotators is including their name or, ideally, a unique researcher identifier, such as ORCiD, as part of annotation metadata. This allows annotation outputs to be directly linked to individuals in a persistent way, to maintain credit for initial annotation in subsequent dataset versions, and to track dataset creation as part of contributions to researcher assessment. Furthermore, data repositories can then use this information to build mechanisms to support the recognition of frequent contributors, those involved in the creation of datasets that are frequently accessed and reused by the scientific community, or other objective measures of the contribution that may be developed in the future.

Widening recognition of the value of open annotated image data requires participation by the whole community. Routes toward this include organising annotation challenges, conducting dataset workshops, and providing training to support the use of annotation standards. Given the importance of annotations for the development of AI methods and, therefore, for the advancement of the bioimaging community, the involvement of funding agencies and data journals in these events is key. These efforts could help incentivize and reward the valuable contributions of annotators, foster a sense of community, and promote and support the creation of high-quality annotations for bioimage analysis.

**Formats: next stop, next generation**

Bioimage annotations that are part of AI-ready datasets come in a plethora of formats, from widely used CSV tables or TIFF segmentation masks to specialised formats such as AnnData for annotated data matrices (Virshup et al. 2021) or YOLO for 2D object detection (Redmon et al. 2016). An important distinction to make when choosing a format is between raster-based and vector/coordinate-based formats. Raster files are pixel-based images, and therefore, raster formats are better to unambiguously assign pixels to objects. Vector files contain mathematically defined shapes, making vector-based formats the better choice to handle sparse and overlapping annotations (especially for very large images, like pathology slides).

Format considerations are one aspect of the overall issue of user accessibility. Since many imaging modalities, such as light-sheet (Stelzer et al. 2021) or volume EM (Peddie et al. 2022), can create very large datasets, providing data in cloud-ready formats that support easy online access to subsets of data without the need for very large downloads is important. To achieve this, formats should also be compatible with community tools for data visualisation and annotation. Some imaging techniques, like confocal microscopy, can generate image sequences in time and space. Therefore, formats that can handle multi-dimensional data will be necessary.

Although it is important to give annotators flexibility when creating their data, there is a need to a) select a manageably small set of common formats that both data generators and consumers can easily read, write and edit and b) work towards ways to combine images and annotations into a single container format. For raster-based segmentations, **OME-Zarr** is a cloud-ready, scalable, and interoperable open format (Moore et al. 2023). Importantly, for AI-ready datasets, annotations such as segmentation masks can be packaged with the image data, allowing provenance tracking. Demonstrating its growing role in providing a common distribution format for the biological imaging community, there are several OME-Zarr tools and libraries for image and annotation visualisation, and tools for format conversion (Moore et al. 2023), however work is still needed for OME-Zarr to support vector-based annotations.

The development of parsers to work with the outputs of commercial platforms for data processing would also improve data accessibility. For example, in the context of spatial omics profiling, the *spatialdata-io* effort (https://spatialdata.scverse.org/projects/io) provides an open-source, community-maintained set of readers from several commercial platforms, that enables the conversion of the data to a modular in-memory representation (Marconato et al. 2024), and the subsequent conversion on disk to OME-Zarr. The process ensures that transformations and spatial information are parsed and saved correctly.

Careful selection of formats for vector-based annotations is recommended. A widely used option to store polygons and shapes is **GeoJSON** (Butler et al. 2016), a JavaScript Object Notation (JSON), multi-language format that supports several well-defined geometrical 2D data primitives such as 'polygon' and 'multipolygon', alongside 'linestring', 'linearring', 'point', 'multipoint'. Furthermore, for applications where performance is a concern, GeoJSON can be readily converted into other formats, such as GeoParquet (https://geoparquet.org/). For 3D vector annotations there are a range of formats describing meshes and point clouds. For example, in the biological domain **EMDB-SFF**, the Segmentation File Format from the Electron Microscopy Data Bank, allows the storage of shape primitives (cones, cuboids, cylinders), surface meshes, and 3D segmentation masks while also supporting structured textual annotation using ontological terms as well as archive accessions.

Additionally, “dataset” formats can aggregate multiple annotations for a set of images. One example is the format from the **COCO** (Common Objects in Context) dataset (Lin et al. 2014), a JSON format that supports multiple annotation types such as segmentations and categorical classes, and is language agnostic. COCO was developed primarily for two-dimensional computer-vision tasks and has been widely adopted by the computer-vision community beyond the specific initial dataset for which it was created. This format is starting to be used by the bioimaging community. However, changes will be required to address specific needs in the field, such as handling segmentations of very large 2D images (e.g. those commonly used in histopathology) or data in three or higher dimensions.

**Improving findability, accessibility and presentation of AI-ready datasets**

Having a central repository for AI-ready bioimage datasets, such as the BIA, will greatly help to make data more accessible. Suggestions to increase findability within data repositories include allowing metadata search, supported by ontologies, and enabling users to find images that are similar to a supplied example. Standardising metadata according to the guidelines described above will be key to enable search. More generally, image archives should establish relationships with scientific journals to explore the possibility of including links to specific datasets in papers. Once shared, making annotated image data discoverable by a range of different data consumers is critical. Progress in the use of AI methods within biological imaging is critically dependent on being able to mix members of different communities, including imaging scientists, research biologists, and AI specialists. Presenting AI-ready data in forms that are understandable by all of these groups is crucial to ensure meaningful interpretation of annotations. API access is important for data integration use cases, as well as for allowing direct data querying, transformation, and model training. A further advantage to this approach is that it supports standardised access to data across multiple different storage systems, without requiring homogeneous data storage.

To guarantee robust quality control, it is essential for datasets to have unique identifiers. Repeated rounds of annotation may be created with newer models, for larger subsets of the data, for different features within the same dataset, or to fix problems with the annotations. Therefore, metadata versioning should be allowed. To ensure that the quality of the annotations is maintained, a system to indicate when a dataset needs updates and to submit proposals for annotation improvements should be implemented. However, for this to work, the community will need to agree on protocols that allow experts to validate the annotations and who the reviewing experts would be in each case. A simpler alternative would be to describe the changes between two versions on the metadata so users can make an informed decision when using a set of annotations.

## Discussion

As funders and scientific publishers strongly endorse open data sharing, and data deposition in repositories becomes a mandatory part of the publication process (“Data Sharing Is the Future” 2023; Kaiser and Brainard 2023), scientific communities need to create guidelines to ensure that the archived data is actually useful (Sever 2023). The MIFA guidelines aim to improve FAIR sharing of annotated bioimages by recommending four guiding principles: standardisation of metadata, incentivising dataset production, reducing the number of data formats used in the field, and increasing data accessibility (Figure 3).

These principles present a high level roadmap towards FAIRer AI image data. Full implementation of the guidelines will require detailed work on schemas and models (particularly for software developers). Here we offer some suggestions as to how individual stakeholders can start adopting the MIFA recommendations:

- **Data producers** can immediately start using the MIFA metadata model to capture key contextual information about their annotations. Supporting toolkits, training, and high-quality examples will help with adoption.
- **Data repositories** should develop deposition pipelines that allow users to include such metadata in their submissions. They can also implement changes to their existing presentation mechanisms according to the MIFA guidelines, for example, to allow data browsing and provide APIs for direct data access.
- **Software developers** can include support for the recommended formats in their development of the workflows, tools, and programming languages used by data producers, annotators, and AI scientists. Additionally, they can support the inclusion of vector annotation formats in OME-Zarr.
- **Funders** can require that annotated datasets that result from funded projects are shared according to the MIFA guidelines.

- **The whole community**, including data repositories, journals, funders, software developers, and scientists can participate in the organisation of events to encourage the production of new high-quality annotated datasets (Uhlmann et al. 2024).

We hope that the adoption of these guidelines by various stakeholders will help accelerate the generation and reuse of AI methods for better analysis of biological images, and all of the downstream benefits to life-sciences research that this will enable. Beyond that, the MIFA guidelines would benefit any annotated dataset of biological images, not only those intended for AI applications. Standardising annotated images would also help the development of classical image analysis algorithms and would permit data aggregation in order to create larger datasets. Many aspects of these guidelines also apply to medical or preclinical imaging. However, given the additional complexity of managing patient-identifiable information, differences in imaging modalities, and existing medical-specific standards, we chose to focus on biological imaging. Aligning the guidelines with parallel initiatives in diagnostic imaging would be useful further work.

Paradigms such as self-supervised learning enable algorithms to learn image features from unannotated training sets, potentially eliminating the need for image annotation (Jing and Tian 2021). However, the progress of self-supervised methods for bioimage analysis is limited by the lack of of large collections of analogous images with consistent, harmonised metadata, which require significant human effort to assemble. Curating such collections, for use in un/self-supervised learning could catalyse the growth of these techniques. For this, a unique metadata set will be required, encompassing the metadata outlined in Supplementary Table 1, excluding the annotations module.

AI models can perform image analysis tasks that would be difficult or impossible to achieve using traditional methods. However, the performance of AI models depends on the quality of the data used to train them. Therefore, providing open, diverse, and high-quality annotated datasets will help unlock the potential of AI and push the state of the art in the bioimage analysis field. To achieve this, it is essential to strengthen the synergy between life scientists and AI developers. By incentivising data producers to openly share expertly annotated datasets with the community, we can accelerate the development of new methods and guide developers towards the new and exciting challenges scientists encounter when analysing their images. This way we will empower researchers in diverse fields with AI tools that will ultimately lead to new discoveries and improved understanding of living systems.

**Figures**

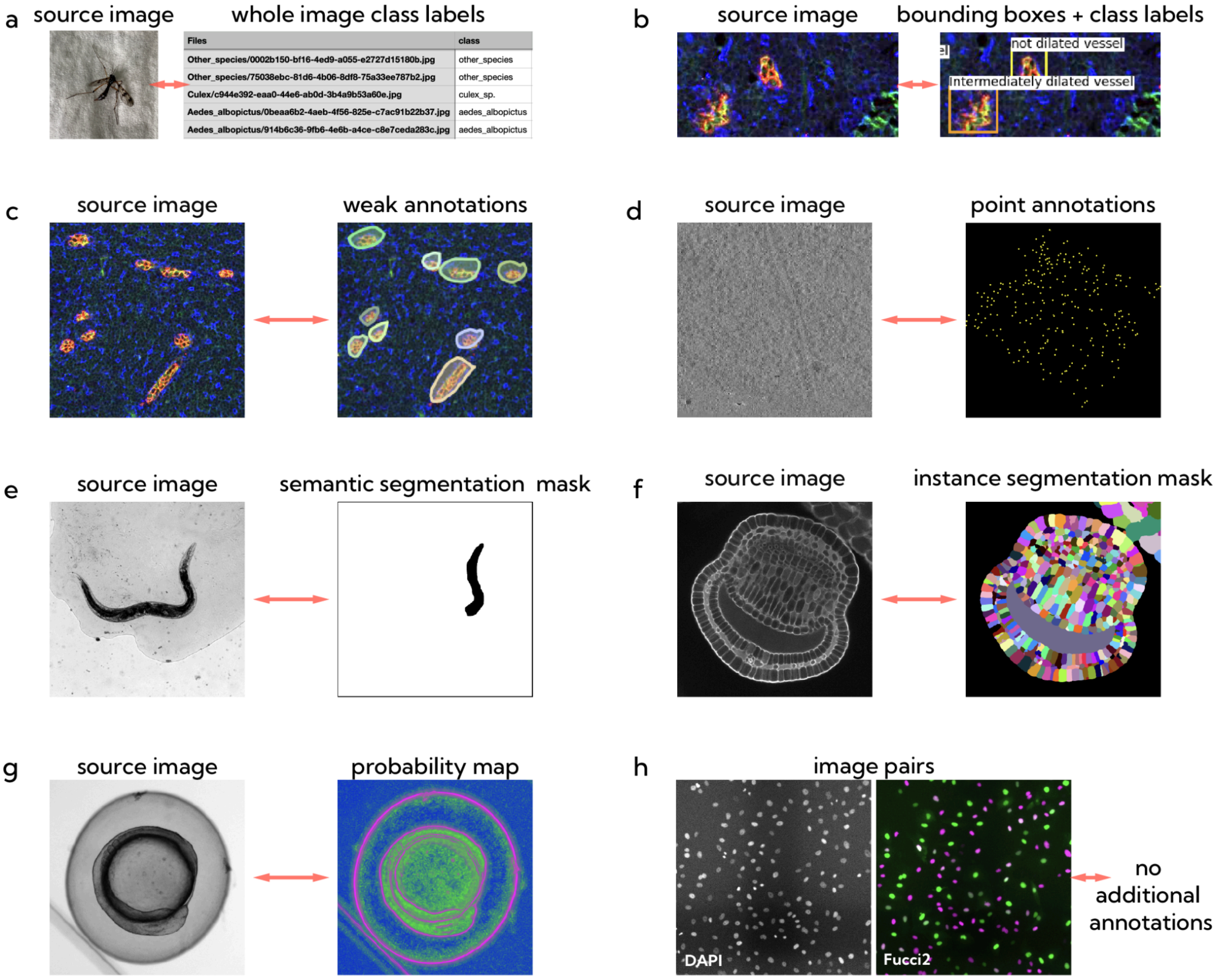

**Figure 1. Diverse annotation types belonging to AI-ready datasets stored at the BioImage Archive and EMPIAR**. **(a)** Whole image class labels designating mosquito genera (accession number: S-BIAD249, http://www.mosquitoalert.com/en/). Unknown image size. **(b)** Bounding boxes and class labels indicating high endothelial venules with different degrees of dilation (accession number: S-BIAD463 (Bekkhus et al. 2021)). Unknown image size. **(c)** Weak manual annotations roughly noting high endothelial venules (accession number: S-BIAD463 (Bekkhus et al. 2021)). Unknown image size. **(d)** Point annotations noting the coordinates of the centres of ribosomes in *S. cerevisiae* (accession number: EMPIAR-11658 (Rangan et al. 2024). Image size, 444.9 x 444.9 Å. **(e)** Semantic segmentation mask of a *C. elegans* head (accession number: S-BIAD300 (Galimov and Yakimovich 2022)). Unknown image size. **(f)** Instance segmentation mask for all the cells in an image of *Utricularia gibba* (accession number: S-BSST734 (Vijayan et al. 2022)). Image size, 208.1 x 208.1 µm **(g)** Probability map indicating the likelihood of each pixel belonging to a boundary (accession number: S-BIAD531 (Jones et al. 2022)). Image size, 1.6 x 1.6 mm **(h)** Pair of images showing the same cells expressing nuclear (left) and cell cycle progression (right) markers. The association between the images is a form of annotation, and these pairs of images alone can be used for supervised training (accession number: S-BSST323 (Rappez et al. 2020)). Unknown image size.

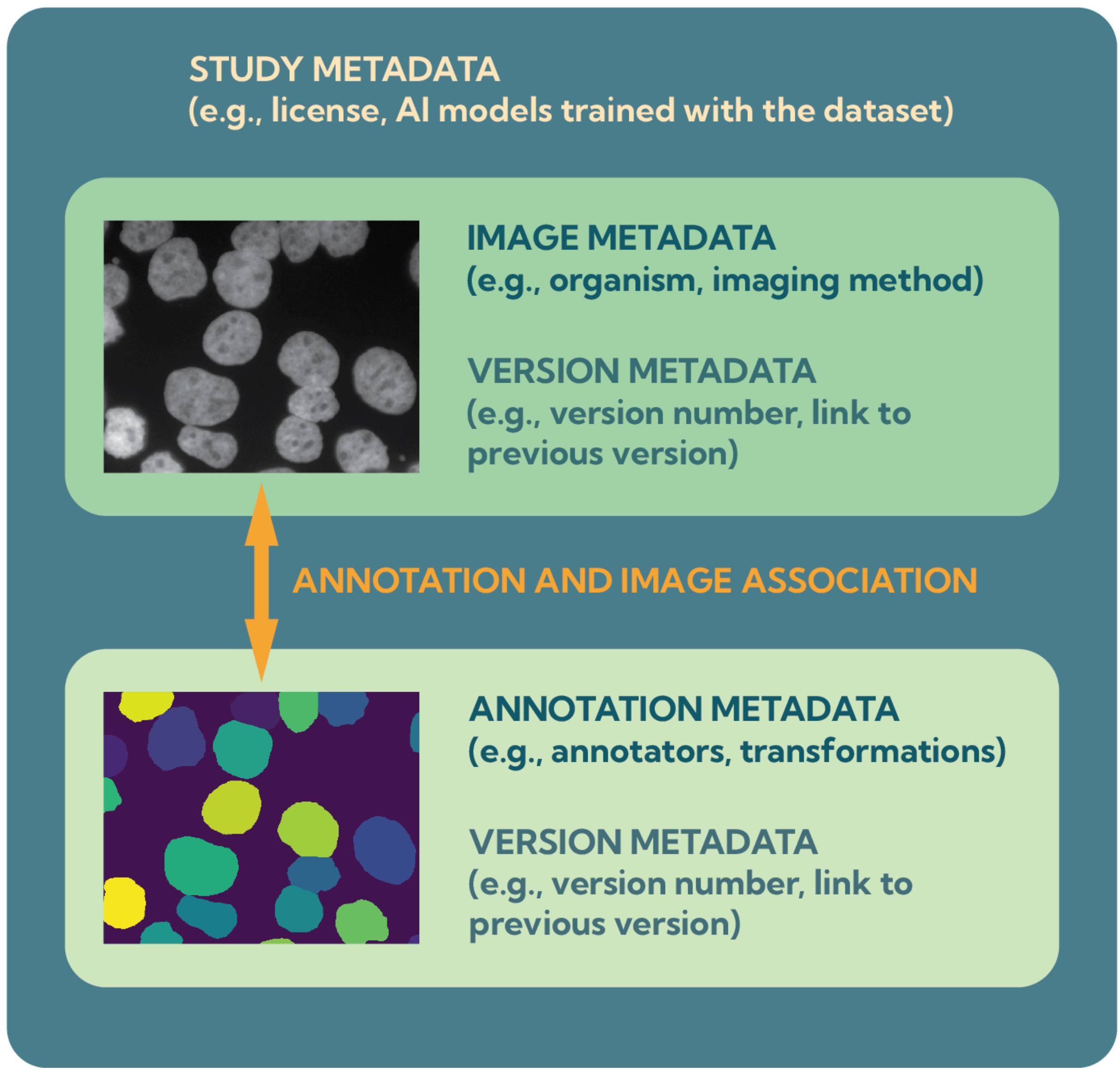

**Figure 2. Metadata modules for AI image datasets.** To make AI-ready datasets reusable, metadata must include general information about the study, the images, the annotations and the version for the image and annotation modules. Furthermore, the association between the annotations and the corresponding images must be clear. Images are from a BIA study with accession number S-BIAD634 (Kromp et al. 2020).

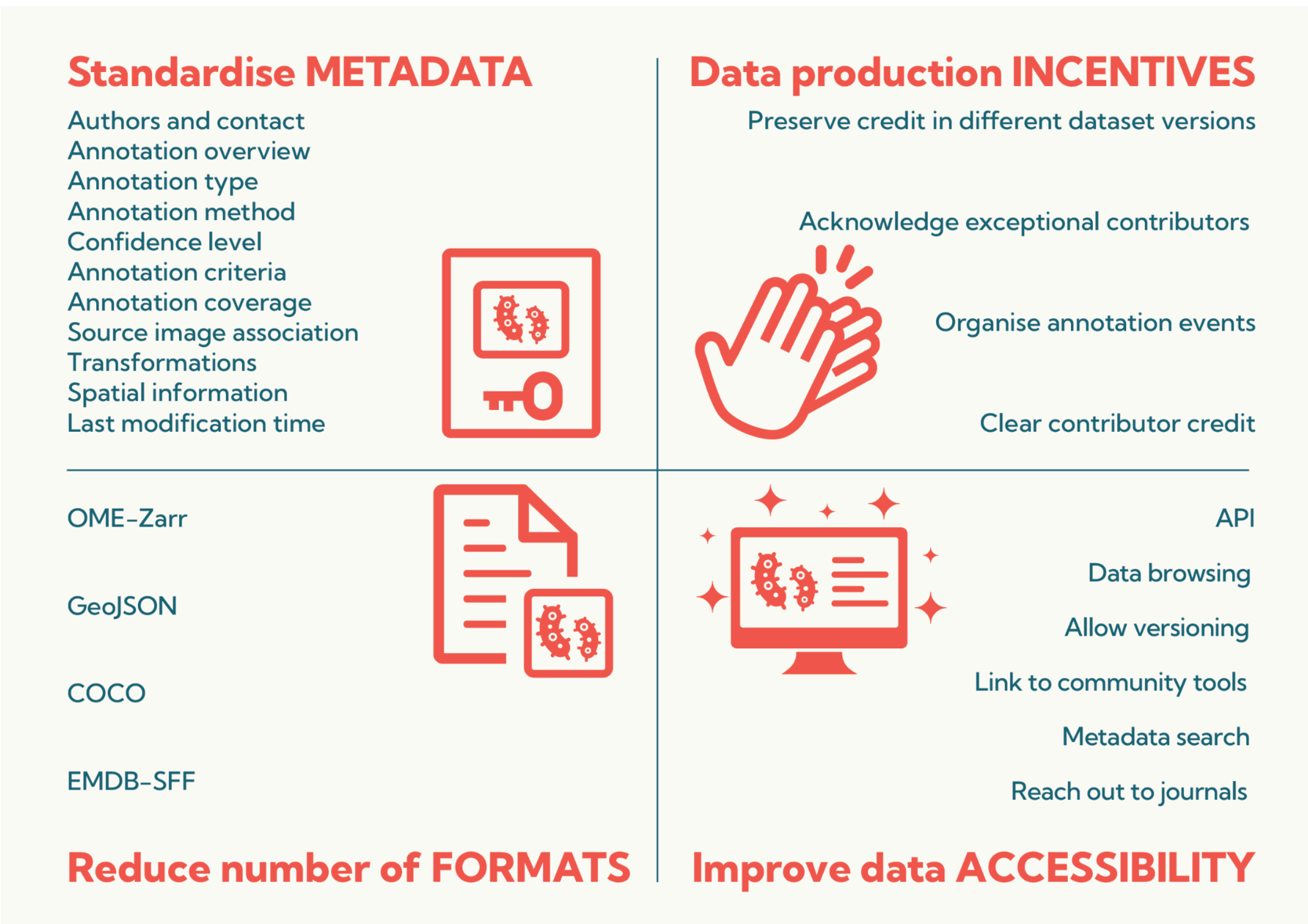

**Figure 3. MIFA (Metadata, Incentives, Formats, and Accessibility) recommendations for FAIR AI data sharing.** Our recommendations can be summarised in four principles: standardising metadata, incentivizing dataset production, reducing format diversity, and increasing data accessibility. (Abbreviations. OME: Open Microscopy environment, JSON: JavaScript Object Notation, COCO: Common Objects in Context, EMDB: Electron Microscopy Data Bank, SFF: Segmentation File Format, API: Application Programming Interface).

**Supplementary materials**

**Supplementary Box 1**: Glossary

**Supplementary Table 1**: Full specification of metadata modules and attributes for AI bioimage datasets with examples

**Data availability**

Data availability is not applicable to this article as no new data were created or analysed for this work. All used images are already publicly accessible, permissibly licensed, and referenced by identifier.

**Acknowledgements**

To improve support for image annotations of AI-related datasets and to develop annotation standards for the community, a workshop was held with 45 community experts from various backgrounds, including data generators, annotators, curators, AI researchers, bioimage analysts, and software developers. The workshop sessions resulted in a series of recommendations on four main topics: Metadata, Incentives, Formats, and Accessibility (MIFA), which are described above. We are grateful to the FAIR AI workshop participants F. Ballllosera, A. Bhardwaj, J.-M. Burel, A. French, M. Hammer, D. Hensen, K. Ho, S. Jasek, I. Kemmer, J. Kriel, A. Iudin, W. Ouyang, A. Papaleo, A. Rupaningal, C. Strambio De Castillia, B.Wester, S. Weyand, and G. Zaki for their insightful inputs and valuable contributions to the discussion.

The workshop was organised in the framework of the AI4Life project, which has received funding from the European Union's Horizon Europe research and innovation programme under grant agreement 101057970. Views and opinions expressed are, however, those of the authors only and do not necessarily reflect those of the European Union. Neither the European Union nor the granting authority can be held responsible for them. This project has been funded in whole or in part with Federal funds from the National Cancer Institute, National Institutes of Health, under Contract No. 75N91019D00024. The content of this publication does not necessarily reflect the views or policies of the Department of Health and Human Services, nor does mention of trade names, commercial products, or organizations imply endorsement by the U.S. Government.

J.M. is supported by the Deutsche Forschungsgemeinschaft (DFG, German Research Foundation; 501864659 as part of NFDI4BIOIMAGE). M.L.J. is supported by the Francis Crick Institute which receives its core funding from Cancer Research UK (CC1076), the UK Medical Research Council (CC1076), and the Wellcome Trust (CC1076). G.J.K. and P.K. were supported by EMBL-EBI and the Wellcome Trust (221371/Z/20/Z). N.N. acknowledges support from the Swedish Research Council (2023-05450), Sigurd and Elsa Goljes Memorial Foundation, IngaBritt och Arne Lundberg foundation, Magnus Bergvall foundation, and Greta and Johan Kock's foundations. V. Ulman was supported by the Ministry of Education, Youth and Sports of the Czech Republic through the e-INFRA CZ (90254). C.T. and V. Ulman. are supported by grant numbers 2020-225265 and 2024-342803, respectively, from the Chan Zuckerberg Initiative DAF, an advised fund of Silicon Valley Community Foundation. P.P-G. is a member of the national infrastructure France-Bioimaging supported by the French national research agency (ANR-24-INBS-0005 FBI BIOGEN).

## References

Bard, Jonathan B. L., and Seung Y. Rhee. 2004. "Ontologies in Biology: Design, Applications and Future Challenges." *Nature Reviews. Genetics* 5 (3): 213–22.

Bekkhus, Tove, Teemu Martikainen, Anna Olofsson, Mathias Franzén Boger, Daniel Vasiliu Bacovia, Fredrik Wärnberg, and Maria H. Ulvmar. 2021. "Remodeling of the Lymph Node High Endothelial Venules Reflects Tumor Invasiveness in Breast Cancer and Is Associated with Dysregulation of Perivascular Stromal Cells." *Cancers* 13 (2). https://doi.org/10.3390/cancers13020211.

Betzig, Eric, George H. Patterson, Rachid Sougrat, O. Wolf Lindwasser, Scott Olenych, Juan S. Bonifacino, Michael W. Davidson, Jennifer Lippincott-Schwartz, and Harald F. Hess. 2006. "Imaging Intracellular Fluorescent Proteins at Nanometer Resolution." *Science* 313 (5793): 1642–45.

Boehm, Ulrike, Glyn Nelson, Claire M. Brown, Steve Bagley, Peter Bajcsy, Johanna Bischof, Aurelien Dauphin, et al. 2021. "QUAREP-LiMi: A Community Endeavor to Advance Quality Assessment and Reproducibility in Light Microscopy." *Nature Methods* 18 (12):

1423–26.
Butler, H., M. Daly, A. Doyle, S. Gillies, S. Hagen, and T. Schaub. 2016. “The GeoJSON Format.” RFC Editor. https://doi.org/10.17487/rfc7946.
Caicedo, Juan C., Allen Goodman, Kyle W. Karhohs, Beth A. Cimini, Jeanelle Ackerman, Marzieh Haghighi, Cherkeng Heng, et al. 2019. “Nucleus Segmentation across Imaging Experiments: The 2018 Data Science Bowl.” *Nature Methods* 16 (12): 1247–53.
“CC0 - Creative Commons.” 2009. Creative Commons. January 22, 2009. https://creativecommons.org/share-your-work/public-domain/cc0/.
Chen, Bi-Chang, Wesley R. Legant, Kai Wang, Lin Shao, Daniel E. Milkie, Michael W. Davidson, Chris Janetopoulos, et al. 2014. “Lattice Light-Sheet Microscopy: Imaging Molecules to Embryos at High Spatiotemporal Resolution.” *Science* 346 (6208): 1257998.
Conrad, Ryan, and Kedar Narayan. 2021. “CEM500K, a Large-Scale Heterogeneous Unlabeled Cellular Electron Microscopy Image Dataset for Deep Learning.” *eLife* 10 (April). https://doi.org/10.7554/eLife.65894.
———. 2023. “Instance Segmentation of Mitochondria in Electron Microscopy Images with a Generalist Deep Learning Model Trained on a Diverse Dataset.” *Cell Systems* 14 (1): 58–71.e5.
“Creative Commons — Attribution 4.0 International — CC BY 4.0.” n.d. Accessed August 3, 2023. https://creativecommons.org/licenses/by/4.0/.
Daetwyler, Stephan, Ulrik Günther, Carl D. Modes, Kyle Harrington, and Jan Huisken. 2019. “Multi-Sample SPIM Image Acquisition, Processing and Analysis of Vascular Growth in Zebrafish.” *Development* 146 (6). https://doi.org/10.1242/dev.173757.
“Data Sharing Is the Future.” 2023. *Nature Methods* 20 (4): 471–471.
Edlund, Christoffer, Timothy R. Jackson, Nabeel Khalid, Nicola Bevan, Timothy Dale, Andreas Dengel, Sheraz Ahmed, Johan Trygg, and Rickard Sjögren. 2021. “LIVECell-A Large-Scale Dataset for Label-Free Live Cell Segmentation.” *Nature Methods* 18 (9): 1038–45.
Galimov, Evgeniy, and Artur Yakimovich. 2022. “A Tandem Segmentation-Classification Approach for the Localization of Morphological Predictors of Lifespan and Motility.” *Aging* 14 (4): 1665–77.
Gupta, Anindya, Philip J. Harrison, Håkan Wieslander, Nicolas Pielawski, Kimmo Kartasalo, Gabriele Partel, Leslie Solorzano, et al. 2019. “Deep Learning in Image Cytometry: A Review.” *Cytometry. Part A: The Journal of the International Society for Analytical Cytology* 95 (4): 366–80.
Hallou, Adrien, Hannah G. Yevick, Bianca Dumitrascu, and Virginie Uhlmann. 2021. “Deep Learning for Bioimage Analysis in Developmental Biology.” *Development* 148 (18): dev199616.
Hartley, Matthew, Gerard J. Kleywegt, Ardan Patwardhan, Ugis Sarkans, Jason R. Swedlow, and Alvis Brazma. 2022. “The BioImage Archive - Building a Home for Life-Sciences Microscopy Data.” *Journal of Molecular Biology* 434 (11): 167505.
Hess, Samuel T., Thanu P. K. Girirajan, and Michael D. Mason. 2006. “Ultra-High Resolution Imaging by Fluorescence Photoactivation Localization Microscopy.” *Biophysical Journal* 91 (11): 4258–72.
Huisken, Jan, Jim Swoger, Filippo Del Bene, Joachim Wittbrodt, and Ernst H. K. Stelzer. 2004. “Optical Sectioning Deep inside Live Embryos by Selective Plane Illumination Microscopy.” *Science* 305 (5686): 1007–9.
Iudin, Andrii, Paul K. Korir, Sriram Somasundharam, Simone Weyand, Cesare Cattavitello, Neli Fonseca, Osman Salih, Gerard J. Kleywegt, and Ardan Patwardhan. 2023. “EMPIAR: The Electron Microscopy Public Image Archive.” *Nucleic Acids Research* 51 (D1): D1503–11.
Jing, Longlong, and Yingli Tian. 2021. “Self-Supervised Visual Feature Learning With Deep Neural Networks: A Survey.” *IEEE Transactions on Pattern Analysis and Machine Intelligence* 43 (11): 4037–58.
Jones, Rebecca A., Matthew J. Renshaw, David J. Barry, and James C. Smith. 2022.

“Automated Staging of Zebrafish Embryos Using Machine Learning.” *Wellcome Open Research* 7 (November):275.
Kaiser, Jocelyn, and Jeffrey Brainard. 2023. “Ready, Set, Share!” *Science* 379 (6630): 322–25.
Kromp, Florian, Eva Bozsaky, Fikret Rifatbegovic, Lukas Fischer, Magdalena Ambros, Maria Berneder, Tamara Weiss, et al. 2020. “An Annotated Fluorescence Image Dataset for Training Nuclear Segmentation Methods.” *Scientific Data* 7 (1): 262.
Laine, Romain F., Ignacio Arganda-Carreras, Ricardo Henriques, and Guillaume Jacquemet. 2021. “Avoiding a Replication Crisis in Deep-Learning-Based Bioimage Analysis.” *Nature Methods* 18 (10): 1136–44.
LeCun, Yann, Yoshua Bengio, and Geoffrey Hinton. 2015. “Deep Learning.” *Nature* 521 (7553): 436–44.
Linkert, Melissa, Curtis T. Rueden, Chris Allan, Jean-Marie Burel, Will Moore, Andrew Patterson, Brian Loranger, et al. 2010. “Metadata Matters: Access to Image Data in the Real World.” *The Journal of Cell Biology* 189 (5): 777–82.
Lin, Tsung-Yi, Michael Maire, Serge Belongie, Lubomir Bourdev, Ross Girshick, James Hays, Pietro Perona, Deva Ramanan, C. Lawrence Zitnick, and Piotr Dollár. 2014. “Microsoft COCO: Common Objects in Context.” http://arxiv.org/abs/1405.0312.
Ljosa, Vebjorn, Katherine L. Sokolnicki, and Anne E. Carpenter. 2012. “Annotated High-Throughput Microscopy Image Sets for Validation.” *Nature Methods* 9 (7): 637.
Marconato, Luca, Giovanni Palla, Kevin A. Yamauchi, Isaac Virshup, Elyas Heidari, Tim Treis, Wouter-Michiel Vierdag, et al. 2024. “SpatialData: An Open and Universal Data Framework for Spatial Omics.” *Nature Methods*, March. https://doi.org/10.1038/s41592-024-02212-x.
Maška, Martin, Vladimír Ulman, Pablo Delgado-Rodriguez, Estibaliz Gómez-de-Mariscal, Tereza Nečasová, Fidel A. Guerrero Peña, Tsang Ing Ren, et al. 2023. “The Cell Tracking Challenge: 10 Years of Objective Benchmarking.” *Nature Methods*, May. https://doi.org/10.1038/s41592-023-01879-y.
McDole, Katie, Léo Guignard, Fernando Amat, Andrew Berger, Grégoire Malandain, Loïc A. Royer, Srinivas C. Turaga, Kristin Branson, and Philipp J. Keller. 2018. “In Toto Imaging and Reconstruction of Post-Implantation Mouse Development at the Single-Cell Level.” *Cell* 175 (3): 859–76.e33.
Megason, Sean G. 2009. “In Toto Imaging of Embryogenesis with Confocal Time-Lapse Microscopy.” *Methods in Molecular Biology* 546:317–32.
Moen, Erick, Dylan Bannon, Takamasa Kudo, William Graf, Markus Covert, and David Van Valen. 2019. “Deep Learning for Cellular Image Analysis.” *Nature Methods* 16 (12): 1233–46.
Moore, Josh, Daniela Basurto-Lozada, Sébastien Besson, John Bogovic, Jordão Bragantini, Eva M. Brown, Jean-Marie Burel, et al. 2023. “OME-Zarr: A Cloud-Optimized Bioimaging File Format with International Community Support.” *Histochemistry and Cell Biology* 160 (3): 223–51.
Nakane, Takanori, Abhay Kotecha, Andrija Sente, Greg McMullan, Simonas Masiulis, Patricia M. G. E. Brown, Ioana T. Grigoras, et al. 2020. “Single-Particle Cryo-EM at Atomic Resolution.” *Nature* 587 (7832): 152–56.
Ouyang, Wei, Fynn Beuttenmueller, Estibaliz Gómez-de-Mariscal, Constantin Pape, Tom Burke, Carlos Garcia-López-de-Haro, Craig Russell, et al. 2022. “BioImage Model Zoo: A Community-Driven Resource for Accessible Deep Learning in BioImage Analysis.” *bioRxiv*. https://doi.org/10.1101/2022.06.07.495102.
Peddie, Christopher J., Christel Genoud, Anna Kreshuk, Kimberly Meechan, Kristina D. Micheva, Kedar Narayan, Constantin Pape, et al. 2022. “Volume Electron Microscopy.” *Nature Reviews Methods Primers* 2 (1): 1–23.
Rangan, Ramya, Ryan Feathers, Sagar Khavnekar, Adam Lerer, Jake D. Johnston, Ron Kelley, Martin Obr, Abhay Kotecha, and Ellen D. Zhong. 2024. “CryoDRGN-ET: Deep Reconstructing Generative Networks for Visualizing Dynamic Biomolecules inside Cells.” *Nature Methods* 21 (8): 1537–45.

Rappez, Luca, Alexander Rakhlin, Angelos Rigopoulos, Sergey Nikolenko, and Theodore Alexandrov. 2020. “DeepCycle Reconstructs a Cyclic Cell Cycle Trajectory from Unsegmented Cell Images Using Convolutional Neural Networks.” *Molecular Systems Biology* 16 (10): e9474.

Redmon, Joseph, Santosh Divvala, Ross Girshick, and Ali Farhadi. 2016. “You Only Look Once: Unified, Real-Time Object Detection.” In *2016 IEEE Conference on Computer Vision and Pattern Recognition (CVPR)*. IEEE. https://doi.org/10.1109/cvpr.2016.91.

Royer, Loïc A., William C. Lemon, Raghav K. Chhetri, Yinan Wan, Michael Coleman, Eugene W. Myers, and Philipp J. Keller. 2016. “Adaptive Light-Sheet Microscopy for Long-Term, High-Resolution Imaging in Living Organisms.” *Nature Biotechnology* 34 (12): 1267–78.

Rust, Michael J., Mark Bates, and Xiaowei Zhuang. 2006. “Sub-Diffraction-Limit Imaging by Stochastic Optical Reconstruction Microscopy (STORM).” *Nature Methods* 3 (10): 793–95.

Rutschi, Corinna, Nicholas Berente, and Frederick Nwanganga. 2023. “Data Sensitivity and Domain Specificity in Reuse of Machine Learning Applications.” *Information Systems Frontiers*, April. https://doi.org/10.1007/s10796-023-10388-4.

Sarkans, Ugis, Wah Chiu, Lucy Collinson, Michele C. Darrow, Jan Ellenberg, David Grunwald, Jean-Karim Hériché, et al. 2021. “REMBI: Recommended Metadata for Biological Images-Enabling Reuse of Microscopy Data in Biology.” *Nature Methods* 18 (12): 1418–22.

Schapiro, Denis, Clarence Yapp, Artem Sokolov, Sheila M. Reynolds, Yu-An Chen, Damir Sudar, Yubin Xie, et al. 2022. “MITI Minimum Information Guidelines for Highly Multiplexed Tissue Images.” *Nature Methods* 19 (3): 262–67.

Schwendy, Mischa, Ronald E. Unger, and Sapun H. Parekh. 2020. “EVICAN-a Balanced Dataset for Algorithm Development in Cell and Nucleus Segmentation.” *Bioinformatics* 36 (12): 3863–70.

Sever, Richard. 2023. “We Need a Plan D.” *Nature Methods* 20 (4): 473–74.

Stelzer, Ernst H. K., Frederic Strobl, Bo-Jui Chang, Friedrich Preusser, Stephan Preibisch, Katie McDole, and Reto Fiolka. 2021. “Light Sheet Fluorescence Microscopy.” *Nature Reviews Methods Primers* 1 (1): 1–25.

Stringer, Carsen, Tim Wang, Michalis Michaelos, and Marius Pachitariu. 2021. “Cellpose: A Generalist Algorithm for Cellular Segmentation.” *Nature Methods* 18 (1): 100–106.

Swedlow, Jason R., Pasi Kankaanpää, Ugis Sarkans, Wojtek Goscinski, Graham Galloway, Leonel Malacrida, Ryan P. Sullivan, et al. 2021. “A Global View of Standards for Open Image Data Formats and Repositories.” *Nature Methods* 18 (12): 1440–46.

Udan, Ryan S., Victor G. Piazza, Chih-Wei Hsu, Anna-Katerina Hadjantonakis, and Mary E. Dickinson. 2014. “Quantitative Imaging of Cell Dynamics in Mouse Embryos Using Light-Sheet Microscopy.” *Development* 141 (22): 4406–14.

Uhlmann, Virginie, Matthew Hartley, Josh Moore, Erin Weisbart, and Assaf Zaritsky. 2024. “Making the Most of Bioimaging Data through Interdisciplinary Interactions.” *Journal of Cell Science* 137 (20). https://doi.org/10.1242/jcs.262139.

Ulman, Vladimír, Martin Maška, Klas E. G. Magnusson, Olaf Ronneberger, Carsten Haubold, Nathalie Harder, Pavel Matula, et al. 2017. “An Objective Comparison of Cell-Tracking Algorithms.” *Nature Methods* 14 (12): 1141–52.

Vijayan, Athul, Soeren Strauss, Rachele Tofanelli, Tejasvinee Atul Mody, Karen Lee, Miltos Tsiantis, Richard S. Smith, and Kay Schneitz. 2022. “The Annotation and Analysis of Complex 3D Plant Organs Using 3DCoordX.” *Plant Physiology* 189 (3): 1278–95.

Villoutreix, Paul. 2021. “What Machine Learning Can Do for Developmental Biology.” *Development* 148 (1): dev188474.

Virshup, Isaac, Sergei Rybakov, Fabian J. Theis, Philipp Angerer, and F. Alexander Wolf. 2021. “Anndata: Annotated Data.” *bioRxiv*. https://doi.org/10.1101/2021.12.16.473007.

Wang, Shidan, Donghan M. Yang, Ruichen Rong, Xiaowei Zhan, and Guanghua Xiao. 2019. “Pathology Image Analysis Using Segmentation Deep Learning Algorithms.” *The American Journal of Pathology* 189 (9): 1686–98.

Wilkinson, Mark D., Michel Dumontier, I. Jsbrand Jan Aalbersberg, Gabrielle Appleton, Myles Axton, Arie Baak, Niklas Blomberg, et al. 2016. “The FAIR Guiding Principles for Scientific Data Management and Stewardship.” *Scientific Data* 3 (March):160018.

Yip, Ka Man, Niels Fischer, Elham Paknia, Ashwin Chari, and Holger Stark. 2020. “Atomic-Resolution Protein Structure Determination by Cryo-EM.” *Nature* 587 (7832): 157–61.

Zhang, Kaiming, Grigore D. Pintilie, Shanshan Li, Michael F. Schmid, and Wah Chiu. 2020. “Resolving Individual Atoms of Protein Complex by Cryo-Electron Microscopy.” *Cell Research* 30 (12): 1136–39.